\documentclass[12pt,preprint]{aastex}

\usepackage{color}

\newcommand{\msun}{\hbox{M$_\sun$}}

\newcommand{\ygu}{{\sc yguaz\'u-A} code}
\newcommand{\com}[1]{{}{ #1}}  

\newcommand{\nii}{\hbox{[N\,{\sc ii}]\,$\lambda$\,6583}}

\slugcomment{To appear in ApJ Letters.}

\righthead{Shaping point- and mirror-symmetric PPNe by orbital motion}
\lefthead{Haro-Corzo et al.}

\begin{document}

\title{Shaping point- and mirror-symmetric proto-planetary nebulae by
  the orbital motion of the central binary system}

\author{Sinhu\'e A. R. Haro-Corzo, Pablo F. Vel\'azquez, Alejandro C. Raga}
\affil{Instituto de Ciencias Nucleares, Universidad Nacional Aut\'onoma de
  M\'exico, Ciudad Universitaria,
  Apartado Postal 70-543, CP 04510, M\'exico D.F., M\'exico.}
\email{haro,pablo,raga@nucleares.unam.mx}

\author{Angels Riera}
\affil{Departament de F\'\i sica i Enginyeria Nuclear, EUETIB, Universitat
Polit\`ecnica de Catalunya, Comte dUrgell 187, E-08036
Barcelona, Spain.\\
Departament dAstronomia i Meteorologia, Universitat de
Barcelona, Av. Diagonal 647, 08028 Barcelona, Spain.\\
Institut d'Estudis Espacials de Catalunya (IEEC), E-08034 Barcelona, Spain.}

\author{Primoz Kajdic}
\affil{Instituto de Geof\'\i sica, Universidad Nacional Aut\'onoma de
  M\'exico, Ciudad Universitaria, M\'exico D.F., M\'exico.}


\begin{abstract}
  
  We present 3D hydrodynamical simulations of a jet launched from the
  secondary star of a binary system inside a proto-planetary nebula.
  The secondary star moves around the primary in a close eccentric
  orbit. From the gasdynamic simulations we compute synthetic \nii\
  emission maps. Different jet axis inclinations with respect to the
  orbital plane, as well as different orientations of the flow with
  respect to the observer are considered. For some parameter
  combinations, we obtain structures that show point- or
  mirror-symmetric morphologies depending on the orientation of the
  flow with respect to the observer. Furthermore, our models can
  explain some of the emission distribution asymmetries that are
  summarized in the classification given by \citet{sokerhadar02}.

\end{abstract}

\keywords{ISM: jets and outflows --- planetary nebulae: general --- stars: AGB
  and post-AGB --- methods: numerical --- binaries: general}

\section{Introduction}

The shell of ionized gas coming from the ejected envelopes of the very
late stage in the death of an intermediate initial mass (M$_*\le
8$~\msun) star is called Planetary Nebula (PN). Since the onset of
high spatial resolution observational facilities, such as the HST,
optical images of PNe have revealed a rich variety of morphologies,
highly collimated structures and small scale features. These nebulae
have been classified according to their large-scale morphologies as:
bipolar, circular, elliptical or irregular. In particular, bipolar PNe
are axially symmetric, having two lobes with an equatorial waist
between them. 

\citet{sokerhadar02} proposed a classification of PNe depending on
their departure from axisymmetry. The departures from axial symmetry
can be point-symmetry or mirror-symmetry, with respect to the
equatorial plane (which separates the two lobes).

In the last decade, the point-symmetric shape of some PNe and
proto-planetary nebulae (PPNe) has been successfully explained in
terms of collimated outflows.  For example, \citet{lee03} carried out
axisymmetrical simulations for modeling the morphology and emission of
the PPN CRL 618. \citet{velazquez04} \citep[also see][]{riera04}, modeled
the multiple shocked structures, emission and kinematics of the PPN Hen
3-1475 in terms of a precessing jet with a periodic velocity
variation, using 3D hydrodynamical simulations. \citet{velazquez07}
modeled the thermal radio-continuum emission of the PN K3-35 as a
continuous, precessing jet. \citet{guerrero08} present an
observational and numerical study of the PN IC 4634, showing that a
variable velocity, precessing jet is the origin of its point-symmetric
morphology. Some features observed in PNe (i.e. low excitation
knots/clumps moving at supersonic velocities) could be produced by
cosmic ``bullets'' \citep[e.g.,][]{raga07,dennis08}.

The presence of a companion to the AGB star is thought to be required
in order to produce jets and point-symmetric morphologies
\citep[][]{balickfrank02}. The presence of binary star system
progenitors of PNe was initially proposed by \citet{bond78}, who
discovered that UU Sge, the central star of PN Abell 63, is an
eclipsing binary.  \citet{fabian79} suggested that the particular
shape of the Helix nebula can also be due to the action of a binary
system.  \citet{livio79} suggested that non-spherical PNe host close
binaries. Furthermore, the survey of \citet{demarco08} shows strong
evidence that at least 10-15\% of PNe are composed of binary systems
with short orbital periods ($< 3$~days). Eventhough this fraction is
not high enough to be consistent with the large fraction of PNe that
apparently have been shaped by binary interactions, this survey is not
able to detect large orbital periods and therefore the binary
fraction of 10-15\% has been considered a lower limit.

Some mechanism invoked to explain the shaping of bipolar PNe
involve the action of jets launched by the primary or secondary star
\citep[e.g.,][]{morris87,morris90,soker92,livio93, soker00}. In
several models, the jets are supposed to form when a main sequence
star (or a white dwarf) accrete material from the AGB (or post-AGB)
star, forms an accretion disk and blows two jets
\citep[e.g.,][]{frank04,frank06,frank07}. Hydrodynamical simulations
have shown that jets (blown by the AGB star or by a companion) can
account for some of the observed morphologies. \citet{garcia04}
carried out 3D numerical simulations of the interaction of the stellar
winds from the components of a binary system. They found that bipolar
morphologies as narrow-waisted nebulae can be obtained after the
interaction of a slow AGB wind, from the primary star, with a
collimated fast stellar wind produced by the secondary.

In this work, we explore the influence of having a binary system in
order to reproduce the morphology and brightness distribution of
bipolar PPNe. We investigate (3D hydrodynamical simulations) if both
point- and mirror- symmetric PPN could be explained in terms of an
interacting jet (source in orbital motion) with an AGB wind.  Also, we
explore if the observed morphology of the PPN is affected by the
orientation of the flow with respect to the observer. The jet is
ejected by the secondary star (a star in the main sequence), which is
on orbital motion around the remnant of the AGB star. For simplicity,
the AGB wind is \com{considered isotropic, i.e., it has a density
distribution given by $\rho_w \propto r^{-2}$ and a constant wind
velocity  (see section\,3)}.
We consider a close eccentric binary system, calling the suggestions
of \citet{soker98}, \citet{soker00} and \citet{soker01}.

\section{Modeling the orbital motion of a binary system}

We investigate the interaction of a jet with the AGB slow wind in a
binary system scenario, where the primary is the remnant AGB star.
Observations show that many PNe seem to have binary sources
\citep{vanwinckel03,zijlstra07,demarco08}, in which the two stars can be
in different evolutionary stages according to their initial masses.

\begin{figure}
\epsscale{.80}
\plotone{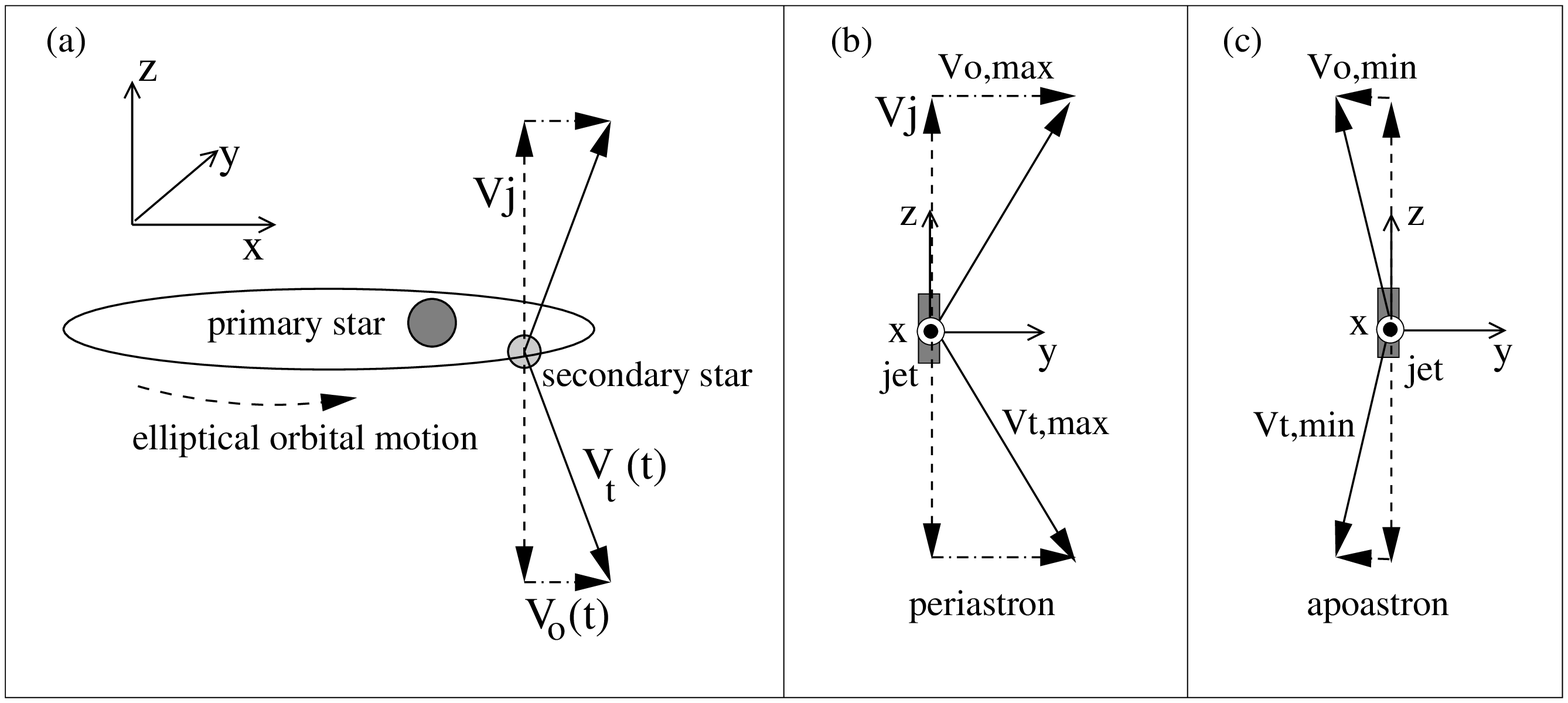}
\caption{\com{(a) Scheme of a vertical jet in orbital motion. The
    center of mass is located in one focus of an elliptical orbit and
    the secondary star orbits in an elliptical path around the center
    of mass (which lies close to the primary star). The
    foci are along the $x$-axis. The $xy$-plane is the orbital
    plane. (b) and (c) are the true scheme employed in our simulations,
    where the jet is located at the origin. The orbital speed is added
    as an additional component of the jet outflow.}}
\label{f1}
\end{figure}

Both stars orbit around the center of mass in
stable elliptical paths. The Keplerian velocity of the reduced mass is
corrected by the factor ${M_p}/(M_s+M_p)$, in order to obtain the
orbital velocity of the secondary star.  The orbital plane (the $xy$-
plane) is such that both foci of the ellipse are located along the
$x$-axis, i.e. both the periastron and apoastron lie along this
axis. With this configuration, the orbital velocity achieves its
maximum magnitude in the periastron, toward the positive $y$-axis (see
\com{panel (a) of} Figure\,\ref{f1}).

\citet{vanwinckel03} gives values for the semi-major axis of the order
of 1 AU, while PNe usually have physical sizes of $\sim 0.1$~pc ($\sim
10^5$~AU). As we are interested in modeling the large scale structure
of PNe, the spatial resolution of our numerical simulations is not
high enough to resolve the scale of the orbital motion. However, our
numerical simulations do resolve the temporal scale of the Keplerian
motion, so that we can include the orbital speed as an additional
component of the outflow velocity. \com{In panels (b) and (c) of
  Figure\,\ref{f1}, we display the actual scheme used, i.e. a fixed jet,
  centered in the computational domain, where the orbital speed is
  added to the jet velocity. Two snapshots are shown for the cases
  where the orbital speed is maximum --at the periastron, panel (b)--
  and minimum --at the apoastron, panel (c)--.}

A detailed exploration of the parameter space (semi-major axis,
eccentricity, etc) is carried out by \citet[in
preparation]{harocorzo09}. In this work, we only show the most relevant
results, choosing \com{$M_p+M_s=1.25$~\msun (with $M_p=1\msun$) and a
semi-major axis of 3.2\,AU (i.e., an orbital period of 5\,years).}

\section{Initial setup for the numerical simulations}

The 3D numerical simulations were carried out with the \ygu\ 
\citep{raga00}. The \ygu\ integrates the gas-dynamical equations together
with a system of rate equations for the atomic/ionic species: HI, HII,
HeI, HeII, HeIII, CII, CIII, CIV, NI, NII, NIII, OI, OII, OIII, OIV,
SII and SIII. With these rate equations, a non-equilibrium cooling
function is computed.  The reaction and cooling rates are described in
detail by \citet{raga02}. The gas-dynamical equations are
integrated with a second order accurate technique (in time and space)
employing the ``flux-vector splitting'' algorithm of \citet{vanleer82}.

Our numerical simulations use a 5-level, binary, adaptive Cartesian grid,
in a computational domain of $256\times 256\times 512$ pixels (at the
highest grid resolution) along the $x$-,$y$-,
and $z$- axes. The size of the computational
domain is of $2\times 10^{17}$cm along the $x$- and $y$-directions, and
$4\times 10^{17}$\,cm along the $z$-direction. 

A jet is injected into a circumstellar medium,
which has been swept up previously by a dense and slow wind from the
AGB central star. Even though the circumstellar medium in bipolar PNe
is aspherical (with higher densities in the equator than along the
bipolar axis), for simplicity and in order to only study the influence
of the binary orbital motion, \com{we have ``turned off'' the AGB influence
imposing on all computational domain the analytical solution of
an isotropic AGB wind}.  Then, the wind density distribution is given
by:

\begin{equation}
\rho_w=\frac{\dot{M}_w}{4\,\pi\,r^2\,v_w}\,,
\label{dagb}
\end{equation}

\noindent where $\dot{M}_w$ is the mass loss rate of the AGB wind, $v_w$ is
the AGB wind terminal velocity, and $r$ is the distance from the primary star.
We assume that the AGB wind is initially neutral, with  $\dot{M}_w=2\times
10^{-6}$\,\msun\,yr$^{-1}$, $v_w=1.5\times 10^{6}$\,cm\,s$^{-1}$
and a temperature T$_w=100$\,K. A decreasing temperature vs. radius
dependence (resulting from the adiabatic expansion of the spherically
diverging wind) is generated as the time-integration progresses.

The jet is injected at the center of the computational domain within a
cylindrical volume with radius $r_j$ and length $l_j$ with values
$r_j=l_j=4\times 10^{15}$\,cm. \com{(which are equivalent to 5 pixels in
the highest resolution grid)}.  The jet temperature was set to
$10^{3}$\,K. The jet velocity is $v_j=1.7\times 10^{7}$\,cm\,s$^{-1}$
and the mean atom\,+\,ion number density is constant with $n_j=5\times
10^{4}$\,cm$^{-3}$. These values correspond to a $\dot{M}_j=1.5\times
10^{-6}$~\msun\, yr$^{-1}$ mass injection, \com{which could seem
high. Due to the limited numerical resolution we are forced to impose
the jet on a region larger than its true size. However, the
total injected mass after $400~\mathrm{yr}$ is $\mathrm{1.2\times
10^{-3}\ \msun\ }$, which looks like reasonable \citep{velazquez07}.

With the parameters listed above, the jet is in the domain of ``strong
jet'' according to the description of \citet{soker00} and
\citet{garcia04}, with a parameter $\chi=\dot{M}_w v_w/ \dot{M}_j
v_j=0.2$. A similar dynamical behaviour could be achieved taking into
account a less dense jet but with higher velocity, keeping constant
the jet ram pressure $\rho_j v^2_j$ and the parameter $\chi
$. However, we prefer to employ a dense jet with a low velocity, being
the last one comparable with the maximum orbital speed, in order to
favor the contribution of orbital motion to the global
morphology of the object. Based in
these initial conditions,}
 we consider six runs labeled as $Ra$i and $Rb$i,
where i=1 corresponds to the case in which the jet axis is
perpendicular to the orbital plane, i=2 to a tilt of 20\degr\ between
the jet axis and the orbital axis, on the $xz$-plane, and i=3 to a
tilt of 20\degr\ on the $yz$-plane (the orbital axis is parallel to
the $z$-axis, and the $x$-axis contains the foci of the elliptical
orbit, see \S 2). Runs $Ra$i have an eccentricity $e=0.73$, and runs
$Rb$i have $e=0.94$. The parameters of the runs are summarized in
Table \ref{t1}.

Finally, from the temperature and atomic/ionic/electronic number
density distributions computed from the numerical time-integration, we
calculate the emission line coefficients of a set of permitted and
forbidden emission lines.  The forbidden line \nii\ was computed by
solving 5-level atom problems, using the parameters of
\citet{mendoza83}.

\section{Results}
\label{sec:res}

We find that many of our jet/AGB wind interaction models have
morphologies that depend on the projection, and that both point- and
mirror- symmetric PNe can be obtained from the same model (depending
on the position of the observer). This is illustrated in Figure
\ref{f2}, where we plot the resulting synthetic \nii\ emission maps at
an integration time of 400~yr. The group of six panels correspond to
the runs $Ra$i. The $yz$- and $xz$-projections of run $Ra1$ show
mirror-symmetric morphology. On the other hand, while the $yz$-
projection of run $Ra2$ has a mirror-symmetric morphology, the
$xz$-projection of this model has a point-symmetric morphology. A
similar switch between point- and mirror-symmetric morphologies is
seen in run $Ra3$.  Also, we analyze the asymmetries of the synthetic
\nii\ emission maps for runs $Ra$i produced by the inclination
(40\degr) of the orbital plane ($xz$-plane) with respect to the plane
of the sky. The intensity maps preserve the mirror- and
point-symmetric morphologies discussed above (for the case in which
the $z$-axis lies on the plane of the sky). Furthermore, we find that
the morphologies of the maps obtained from runs $Rb$i are similar to
the maps from $Ra$i, although in general, we note the presence of
small-scale internal structures which are brighter for runs $Rb$i than
runs $Ra$i. Those are because of the larger eccentricity of the $Rb$i
runs. The velocity variability of the jet (resulting from the orbital
motion), produces these internal substructures in both lobes of the
$Ra$i and $Rb$i maps. In order to see the influence of the increasing
eccentricity, in the figure\,\ref{f3} we display the substraction of
the synthetic \nii\ emission maps corresponding to $Rb$i- $Ra$i runs.

Finally, we have obtained the \nii\ luminosities for the top and
bottom lobes for all of the computed maps.  Table \ref{t2} gives the
\nii\ luminosities for both lobes and all runs. In general, we note
that runs $Rb$i have larger \nii\ luminosities than runs $Rb$i (of the
order of a 10\%), as a result of the more eccentric orbital motion of
runs $Rb$i.  The exception is the run $Rb$3 because the total
luminosity measured on the top lobe is the same of run $Ra$3, while
the bottom lobe has less total luminosity. This is due to the gas
material of top lobe has larger velocities than the bottom lobe
gas. In the top lobe, the orbital motion speed is added to the jet
velocity, while the opposite occurs for the bottom lobe. For the cases
where the resulting shape has mirror-symmetry, the luminosities from
the right and left sides of the lobes have also been
computed. Comparison of total luminosities between both (top and
bottom) lobes show a similar value for runs $Ra1$, $Ra2$, $Rb1$ and
$Rb2$.  However, analyzing the $xz$- and $yz$- projection cases, an
asymmetry in the brightness distribution is observed for the
$yz$-projection, in general, with the right side being more luminous,
consequence of the orbital motion.  In runs $Ra$3 and $Rb$3 we obtain
appreciable \nii\ luminosity asymmetries between the top and bottom
lobes (see figure \ref{f3}).

Point- and mirror-symmetric morphologies are simultaneously obtained
if the jet axis has a certain inclination with respect to the orbital
plane. Interestingly, the classification of a PN belonging to one or
other group seems to depend on the point of view of the observer or on
the projection. There are significant differences in the total
luminosities of both lobes between runs $Ra2$ and $Ra3$ and similar
behavior is observed for runs $Rb2$ and $Rb3$. Run $Rb2$ produces top
and bottom lobes with similar total luminosities, while run $Rb3$
generates a luminosity difference of at least 36\% between the top and
bottom lobes.  This is due to the fact that the jet axis is tilted
20\degr\ on the $yz$-plane in the case of run $Ra3$. In this case, the
jet inclination contributes to the elliptical motion, being its effect
larger for the top lobe when the secondary star passes by the
periastron, where the orbital motion velocity is maximum. The total
jet velocity (the intrinsic constant jet velocity $\sim 1.7\times
10^{7}$\,cm\,s$^{-1}$ plus the eccentric orbital motion velocity) is
also maximum at the periastron, producing strong shocks and increasing
the emission for the top lobe (the opposite occurs for the bottom
lobe).

\section{Discussion and Conclusions}

In this work, we carried out 3D hydrodynamical simulations with
the \ygu\ in order to explore the binarity mechanism for shaping
bipolar, point- and mirror-symmetric PN. We consider a
collimated outflow (jet) launched by the secondary star of a binary
system, located at the center of a PN. The primary star is the source
of the AGB wind, in which the jet is embedded.
The two stars have a close, very eccentric
orbit, resulting in a maximum orbital velocity of the order of the jet
velocity. Because of this, the orbital velocity has a strong effect on
the jet dynamics, and strongly modifies the emission line maps predicted
from the computed flows.

\begin{figure}
\epsscale{1.}
\plotone{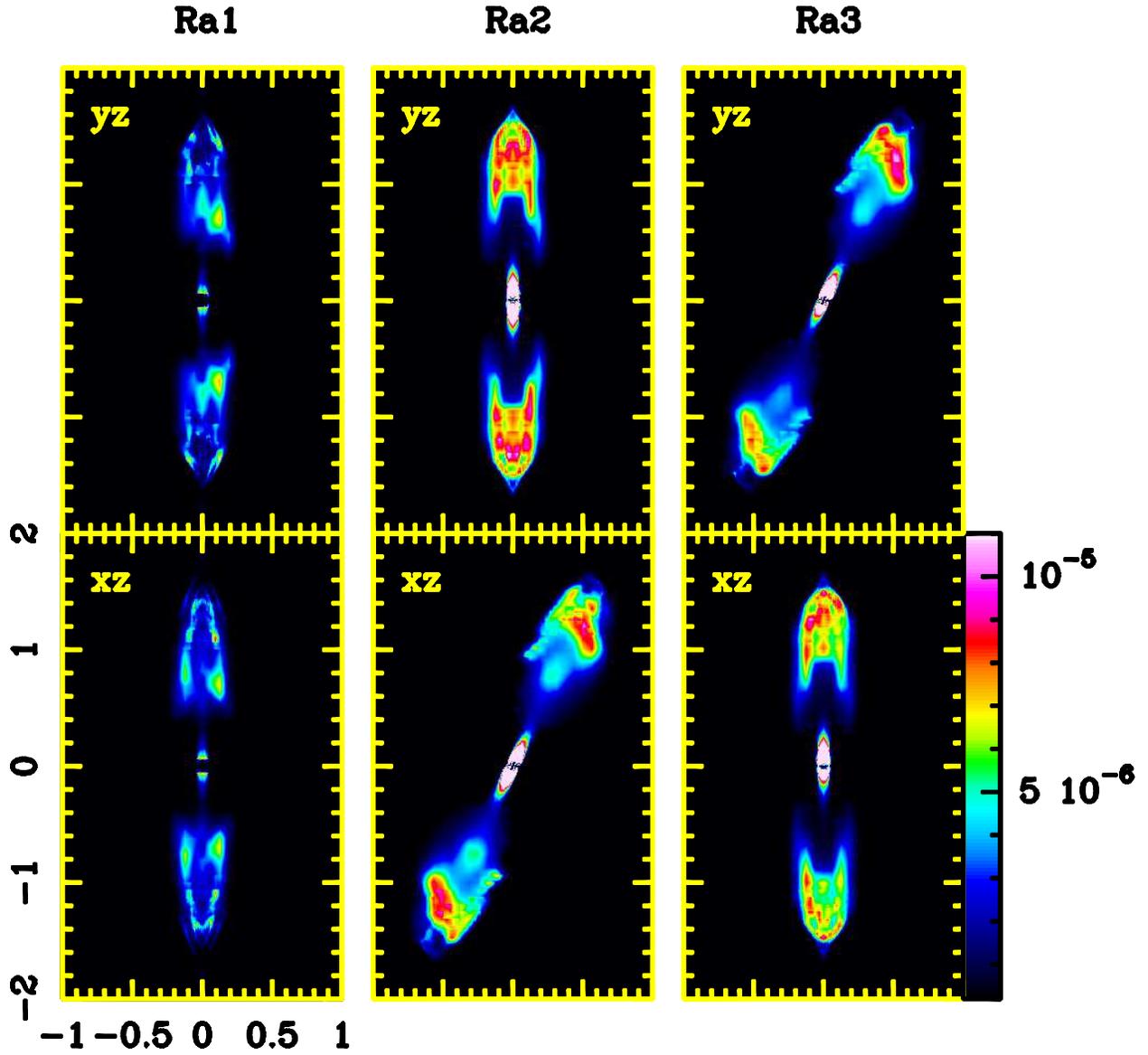}
\caption{ Synthetic \nii\ emission
  \com{($\mathrm{erg\ s^{-1}\ cm^{-2}\ sr^{-1}}$) at 400\,yr for runs
    $Ra1$, $Ra2$ and $Ra3$. The emission is on log gray-scale. Axes
    are in units of $10^{17}$~cm. [See electronic edition for a color version.]}}
\label{f2}
\end{figure}

\begin{figure}
\epsscale{1}
\plotone{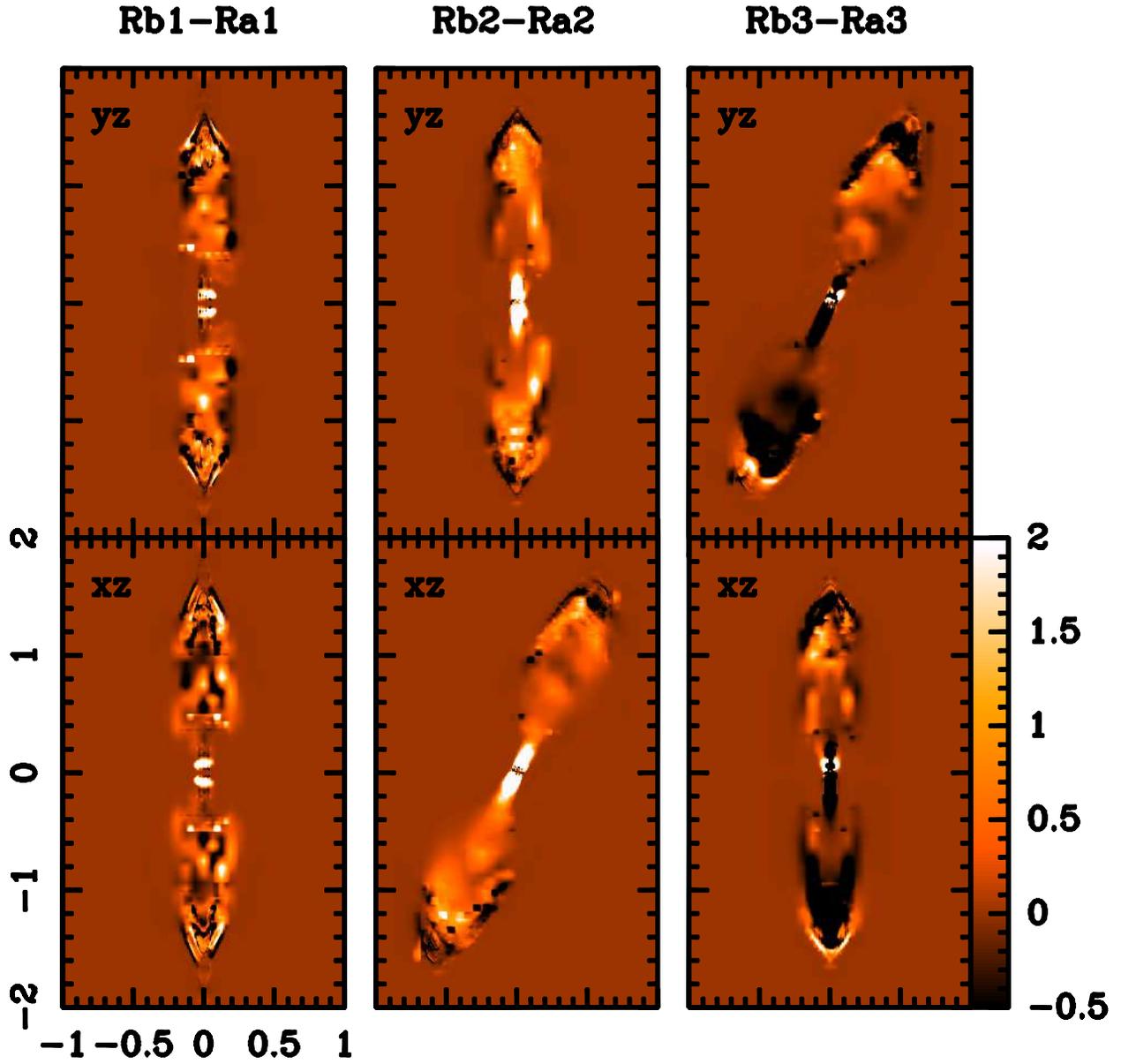}
\caption{Synthetic \nii\, emission \com{at 400\,yr. for the substraction
  $Rb$i- $Ra$i.  Same scale as figure\,\ref{f2}. [See electronic
    edition for a color version.]}}
\label{f3}
\end{figure}

We have generated \nii\ emission maps considering different inclinations
of the jet axis with respect to the orbital plane and also different
tilts between the orbital axis and the line of sight.
We find that the predicted maps of some runs
exhibit (see figure \ref{f2}) either point- or
mirror-symmetric substructures depending on the orientation
of the flow with respect to the observer. Also, the maps exhibit
substructures with different luminosities for the two lobes and
with side-to-side asymmetries within the same lobe.

In this sense, \citet{sokerhadar02} built a classification of PNe
based on their departure from axisymmetry considering both
morphological and brightness distribution criteria.  From Figure
\ref{f2} and table \ref{t2}, we note that runs $Ra3$ and $Rb3$ would
correspond to class 2 (i.e. unequal intensity of both sides) of
\citet{sokerhadar02}, because top lobes are brighter than bottom ones.
Into this class could be also included maps of runs $Ra1$ and $Ra2$,
for example, because in their $yz$-projection display an asymmetry in
the right and left brightness distribution. The $yz$-projections of
$Ra1$ and $Rb1$ could be also included in the class 4 (i.e. the jets
are bent to the same side, therefore there is a mirror-symmetry),
because the small-scale structure observed on the right side of
both lobes.

From our study we conclude that a jet ejected from a companion to the
AGB star (which gives rise to a PN) in a highly eccentric orbit leads
to morphologies with either point- or mirror-symmetries (in some cases
both, depending on the inclination to the observer). Also, the
bipolar nebular structure resulting from the jet/AGB stellar wind
interaction has brightness asymmetries with contrasts similar to the
asymmetries observed in some PNe \citep{sokerhadar02}. 
\com{We must note that, except for the
  eccentricity, the orbital parameters we used are commonly found in
  last evolution state of low mass and small post-AGB stars,
  which can be a close binaries without a common envelope
  \citep{vanwinckel03, frankowski04, vanwinckel09,demarco09}.
The eccentricities we used are uncommon for post-AGB stars, however,
there are reports of highly eccentric orbits in other binary systems,
such as Ba stars \citep{vanwinckel03,jorissen98}.
 \citet{demarco09} argued that for post-AGB stars that have been
  monitored spectroscopically, low eccentricities have been found for
  systems with intermediate periods (between 100 to 1500 days), but 
 for periods $> 1$ yr the eccentricities can be close to unity).
A larger sample of post-AGB in binary systems is needed to assess the
feasibility of our model.}

It is interesting that we can generate models which produce either
point- or mirror-symmetric emission line maps, depending only on the
orientation of the flow with respect to the observer. For other
choices of flow parameters, however, it is easy to obtain models which
produce point-symmetric structures regardless of the orientation of
the flow with respect to the observer \citep[e. g., by having a
precessing jet from a source with low orbital
velocity,][]{masciadri02}. Therefore, while a large range of
parameters will produce point-symmetric nebulae, we find that it is
necessary to have both an appropriate parameter combination and an
appropriate orientation of the flow (with respect to the observer) in
order to produce mirror-symmetric structures. This result might be
consistent with the fact that many more bipolar PNe are observed to
have point-symmetric structures (as opposed to mirror-symmetric
structures).  In order to decide whether this is indeed the case, a
more detailed study limiting the parameter space relevant for nebulae
resulting from jet/AGB wind interactions will have to be made, as well
as a study of the statistics of the resulting flow morphologies
observed at random orientations.

\acknowledgments

Authors acknowledge anonymous referee for her/his very useful comments
and suggestions.
We thank Mart\'in Guerrero and Luis F. Miranda for many clarifying and
useful comments. The authors acknowledge support from grants,
CONACyT 46828-F, and DGAPA IN119709. The work of
A.R. was supported by the MICINN grant AYA2008-06189-C03 and
AYA2008-04211-C02-01 (co-funded with FEDER funds). We also thank Enrique
Palacios,
Antonio Ram\'\i rez and Mart\'\i n Cruz for the assistance provided, and Alejandro Esquivel
for reading this manuscript.


\clearpage

\clearpage

\begin{table}
  \caption{Run parameters. \label{t1} }

  \begin{tabular}{lcccccc}
            \tableline \noalign{\smallskip}

 run    &ecc&      angle(\degr) &plane     \\

\tableline
            \noalign{\smallskip}

Ra1&0.73&    0       &       -       \\
Ra2&0.73&    20      &       xz      \\
Ra3&0.73&    20      &       yz      \\
\\
Rb1&0.94&    0       &       -       \\
Rb2&0.94&    20      &       xz      \\
Rb3&0.94&    20      &       yz      \\

            \tableline
            \noalign{\smallskip}
  \end{tabular}
\end{table}

\begin{table}
  \caption{Luminosity of top and bottom lobes L/L$_{\sun}
  (10^{-5})$. Each lobe has been parted as left and right side.  We
  ignore the central emission (20 pixels from each lobe). \label{t2} }
  \begin{tabular}{lcp{0.1in}p{0.1in}cp{0.1in}p{0.1in}ccp{0.1in}p{0.1in}cp{0.in}p{0.1in}p{0.1in}c}
            \tableline
            \noalign{\smallskip}
   Run & Bottom&&xz&&&yz& & Top&&xz&&&yz& &  \\
   &total&left&&right&left&&right&total&left&&right&left&&right\\
            \tableline
            \noalign{\smallskip}

 Ra1&3.4&1.7&&1.7&1.3&&2.1&3.4&1.7&&1.7&1.3&&2.1  \\
 Ra2&6.9&-&&-&3.2&&3.7&6.9&-&&-&3.2&&3.7  \\
 Ra3&5.6&2.9&&2.7&-&&-&6.9&3.6&&3.3&-&&-  \\
\\
 Rb1&3.8&1.9&&1.9&1.5&&2.3&3.7&1.8&&1.9&1.5&&2.2  \\
 Rb2&7.6&-&&-&3.4&&4.2&7.4&-&&-&3.3&&4.1  \\
 Rb3&5.&2.6&&2.4&-&&-&6.8&3.6&&3.2&-&&-  \\

            \tableline
            \noalign{\smallskip}
  \end{tabular}
\end{table}

\clearpage

\end{document}